\documentclass[a4paper,11pt]{article}
\usepackage{jcappub} 
\usepackage{lineno}
\usepackage{amssymb, amsmath}
\usepackage{tabularx}
\usepackage{graphicx}	
\usepackage{booktabs}
\usepackage{amsmath, mathrsfs}	
\usepackage{subcaption}
\usepackage{float}

	\newcommand{\bq}{\begin{equation}}
	\newcommand{\eq}{\end{equation}}
	\newcommand{\bqn}{\begin{eqnarray}}
	\newcommand{\eqn}{\end{eqnarray}}
	\newcommand{\nb}{\nonumber}
	\newcommand{\lb}{\label}

	\newcommand{\CQG}{Class. Quantum Grav.}

\arxivnumber{1234.56789} 
\title{\boldmath Shadows of Loop Quantum Black Holes: Semi-analytical Simulations of Loop Quantum Gravity Effects on Sagittarius~A$^*$ and M\,87$^*$}

\author[a]{Hong-Xuan Jiang}
\affiliation[a]{Tsung-Dao Lee Institute, Shanghai Jiao Tong University, Shengrong Road 520, Shanghai, 201210, China}
\author[b,c,d]{Cheng Liu}
\affiliation[b]{School of Physics and Astronomy, Shanghai Jiao Tong University, 800 Dongchuan Road, Shanghai, 200240, China}
\affiliation[c]{Shanghai Frontiers Science Center for Gravitational Wave Detection, 800 Dongchuan Road, Shanghai 200240, China}
\affiliation[d]{United Center for Gravitational Wave Physics (UCGWP),  Zhejiang University of Technology, Hangzhou, 310023, China}
\author[a]{Indu K. Dihingia}

\author[a,b]{Yosuke Mizuno}
\author[b,c]{Haiguang Xu}
\author[d,e]{Tao Zhu}
\affiliation[e]{Institute for Theoretical Physics \& Cosmology, Zhejiang University of Technology, Hangzhou, 310023, China}
\author[d,e]{Qiang Wu}

\emailAdd{mizuno@sjtu.edu.cn, zhut05@zjut.edu.cn, hongxuan\_jiang@sjtu.edu.cn}

\abstract{
In this study, we delve into the observational implications of rotating Loop Quantum Black Holes (LQBHs) within an astrophysical framework. We employ semi-analytical General Relativistic Radiative Transfer (GRRT) computations to study the emission from the accretion flow around LQBHs. Our findings indicate that the increase of Loop Quantum Gravity (LQG) effects results in an enlargement of the rings from LQBHs, thereby causing a more circular polarization pattern in the shadow images.
We make comparisons with the Event Horizon Telescope (EHT) observations of Sgr\,A$^*$ and M\,87$^*$, which enable us to determine an upper limit for the polymetric function $P$ in LQG. The upper limit for Sgr\,A$^*$ is $0.2$, while for M\,87$^*$ it is $0.07$. Both black holes exhibit a preference for a relatively high spin ($a\gtrsim0.5$ for Sgr\,A$^*$ and $0.5\lesssim a \lesssim 0.7$ for M\,87$^*$). The constraints for Sgr\,A$^*$ are based on black hole spin and ring diameter, whereas for M\,87$^*$, the constraints are further tightened by the polarimetric pattern. In essence, our simulations provide observational constraints on the effect of LQG in supermassive black holes (SMBH), providing the most consistent comparison with observation.
}
\begin{document}
\maketitle
\flushbottom

\section{Introduction} \label{sec:intro}

General relativity, since its birth in 1915, has withstood the most rigorous and precise experimental tests, such as the classical solar system test \citep{Will:2014kxa}, pulsar test \citep{Hulse:1974eb}, and more recently, the first direct detection of gravitational waves \citep{LIGOScientific:2016aoc} and the first image observations of black holes \citep{EventHorizonTelescope:2019dse}, establishing its position as the standard theory for describing the laws of gravity. Despite being the most successful theory of gravity since its inception, general relativity (GR) encounters challenges both theoretically (e.g., understanding singularity, quantization of gravity, etc.) and observationally (e.g., understanding dark matter, dark energy, etc.). Specifically, Einstein's GR does not incorporate any quantum principles, and it remains an open question: ``How to unify GR and quantum mechanics?'' \citep{Ng:2003jk, Adler:2010wf}. GR also inevitably results in singularities both at the beginning of the universe \citep{Borde:1993xh, Borde:2001nh} and at the center of black hole spacetimes \citep{Hawking:1973uf}, where our known physics laws break down completely.

One of the most effective ways to address these anomalies is to propose various classical modified theories of gravity or quantum gravity. Therefore, it is essential to test the modified theories beyond GR to confirm the final theory of gravity. The loop quantum gravity (LQG) elegantly resolves both the classical Big Bang and black hole singularities. Recently, a self-dual, regular, and static spacetime metric is obtained in the mini-superspace approach based on the polymerization procedure in LQG \cite{Modesto:2008im, Yang:2023gas}. This self-dual spacetime is regular and free of any spacetime curvature singularity. In this spacetime, the effects of LQG are characterized by two parameters, i.e., the minimal area and the Barbero-Immirzi parameter. It is also shown that this spacetime is self-dual in the sense of T-duality \citep[see][for detail]{Modesto:2009ve, Sahu:2015dea}. Over the past ten years or so, many studies have been done on black holes in LQG using various quantization methods. These black holes have been found to have excellent properties and not have any singularities at the origin \cite{Ashtekar:2018cay, Ashtekar:2018lag, Bojowald:2018xxu, ABP19, ADL20, Perez17, Han:2023wxg, Rovelli18, BMM18, Ashtekar20, Gan:2020dkb}. 

A key question has emerged regarding whether LQG effects in the loop quantum spacetime can produce any observational signatures for current and future experiments, allowing us to \textbf{test the effect of LQG}. This has spurred a surge of studies over the past decades, drawing from various experiments and observations \citep[e.g.,][]{Alesci:2011wn, Chen:2011zzi, Dasgupta:2012nk, Hossenfelder:2012tc, Barrau:2014yka, Cruz:2015bcj, add1, add2, Cruz:2020emz, Santos:2021wsw, Liu:2020ola, Sahu:2015dea, Zhu:2020tcf, Virbhadra:2022iiy, Yan:2022fkr, Tu:2023xab,Papanikolaou:2023crz}. In \cite{Liu:2020ola}, LQG effects on the shadow of the rotating black hole are discussed in detail. They also subsequently discussed the observational implications of their results in the latest Event Horizon Telescope (EHT) observation of the supermassive black hole M87. Moreover, the gravitational lensing in the self-dual spacetime is studied, and the polymeric function from LQG is constrained by using the Very-Long-Baseline Interferometry (VLBI) data of the solar gravitational deflection of radio waves \citep{Sahu:2015dea}. Recently, the solar system test and celestial mechanics test by S2 orbit around Sgr\,A$^*$ of the self-dual spacetime are considered in \cite{Zhu:2020tcf} and \cite{Yan:2023vdg} respectively. They derived the observational constraints of the polymeric function $P$ of LQG. It is interesting to note that the phenomenological studies of other types of loop quantum black holes have also been widely explored \cite{Liu:2021djf, Daghigh:2020fmw, Bouhmadi-Lopez:2020oia, Fu:2021fxn, Brahma:2020eos}. However, there are still very few studies performed to extract the observable effects of rotating LQBH in a realistic astrophysical environment.

Recently, the EHT collaboration released the first direct image of a black hole in human history, which captured a horizon scale image of a supermassive black hole at the center of a neighboring elliptical galaxy M87 \cite{EventHorizonTelescope:2019dse, Akiyama:2019brx, Akiyama:2019eap, Akiyama:2019bqs, Akiyama:2019fyp, Akiyama:2019sww}.
They found that the diameter of the center black hole shadow is $(42\pm 3) \; {\rm \mu as}$ with a deviation $\lesssim 10 \%$ from circularity from this image. This leads to a measurement of the center mass, $M=(6.5\pm 0.7)\times 10^9 M_{\odot}$  \cite{EventHorizonTelescope:2019dse}. These results are in good agreement with the predictions of GR by assuming the geometry of black holes is described by the Kerr metric.
Shortly thereafter, the EHT did horizon scale observation of Sgr\,A$^*$, which is the highest angular resolution observation so far \citep{EventHorizonTelescope:2022wkp,
EventHorizonTelescope:2022apq,
EventHorizonTelescope:2022wok,
EventHorizonTelescope:2022exc,
EventHorizonTelescope:2022urf,
EventHorizonTelescope:2022xqj}. Sgr\,A$^*$ shows complicated features that remain unsettled and yet active subjects of study. The rapid intrinsic variation puts a great challenge to do a detailed observation of it \citep{EventHorizonTelescope:2022exc}. Flares are observed at near-infrared and X-ray \citep[e.g.,][]{Genzel:2003as,
Mossoux:2020ddc}, which is also one of the major frontiers of investigation. Multiple astrophysical models have been proposed to explain the observed features of Sgr\,A$^*$ \citep{Jiang:2023ygs, Liska:2017alm,
Ressler:2020voz,
Ripperda:2021zpn}.
All these studies, though, are predicated on the idea that the Kerr metric provides the geometry of the black hole. There has not been a significant deviation from GR detected yet in the constraints of EHT observation \citep{EventHorizonTelescope:2022xqj}. Nevertheless, the alternative theory of gravity cannot be ruled out \citep{Mizuno:2018lxz, EventHorizonTelescope:2021dqv}, and more study is required. The observations by EHT can still be used to probe the environment around the black hole and distinguish or constrain black holes in different theories of gravity \citep{Tsukamoto:2014tja, Tsukamoto:2017fxq, Chael:2018oym, Kumar:2023jgh, Mizuno:2018lxz, Roder:2023oqa,Psaltis2020, Akiyama:2019eap,EventHorizonTelescope:2022xqj}.

In this paper, we carefully investigate the possible observational effects of rotating LQBH in the astrophysical context by performing semi-analytical GRRT calculations of accretion flow around black holes. Our GRRT calculation is implemented with the LQBH metric and computed at $230\,\rm GHz$. We finally compare our simulation results with existing observations of Sgr\,A$^*$ and M87\,$^*$ to constrain the parameters from LQG.

This manuscript is structured in the following manner: Section 2 describes the foundational concepts of loop quantum black holes. In Section 3, we discuss the effects of LQG on GRRT results. In Section 4, we present the constraint from the EHT observation. Finally, a summary and discussion of our work are presented in Section 5.

\section{A Brief Introduction to LQBH spacetime}\label{sec:LQBH}

The loop quantum spacetime arises from the symmetry-reduced model of LQG, which corresponds to a homogeneous spacetime. Such spacetime has been shown to be geodesically complete and free from any spacetime curvature singularity at the center of the black holes.
The LQBH metric in the Boyer-Lindquist coordinates is given by \cite{Liu:2020ola},
	\bqn
	d s^2 &=& \frac{\mathscr{H}}{\Sigma} \left[\frac{\Delta}{\Sigma}(dt-a\sin^2\theta d\phi)^2-\frac{\Sigma}{\Delta}dr^2-\Sigma d\theta^2-\frac{\sin^2\theta}{\Sigma}(a dt-(k^2+a^2)d\phi)^2 \right],  \lb{mmm} \label{metric}
	\eqn
with
\begin{eqnarray}
\Delta(r)&=& \frac{(r-r_+)(r-r_-)r^2}{(r+r_*)^2}+a^2, \label{Delta}\\
\Sigma(r)&=&k^2(r)+a^2\cos^2\theta,\\
k^2(r)&=& \frac{r^4+a^2_0}{(r+r_*)^2},
\end{eqnarray}
where $a$ is the specific angular momentum (spin parameter) of the black hole. The factor $\mathscr{H}$ in the metric (i.e., in Eq.~\ref{metric}) is a regular function that satisfies $\lim\limits_{a \to 0} \mathscr{H}=r^2+a_0^2/r^2$. Accordingly, the coefficient $\mathscr{H}/\Sigma$ is a normalization factor that does not affect the intrinsic geometry of spacetime. It is easy to verify that when $a=0$, we recover the non-rotating self-dual solution as given by \cite{Zhu:2020tcf}. In addition, $r_+=2 G_\text{LQG} M/(1+P)^2$ and $r_{-} = 2G_\text{LQG} M P^2/(1+P)^2$ are the two horizons for the corresponding spherical LQBH (i.e., the spin $a=0$ case), and $r_{*}= \sqrt{r_+ r_-} = 2G_\text{LQG} MP/(1+P)^2$ with $G_\text{LQG}$ representing the gravitational constant in the LQBH, $M$ denoting the ADM ({Arnowitt-Deser-Misner}) mass of the solution, and $P$ is the polymeric function, which is given by
\bqn
P \equiv \frac{\sqrt{1+\epsilon^2}-1}{\sqrt{1+\epsilon^2}+1},   \label{P_epsilon}
\eqn 
where $\epsilon$ denotes a product of the Immirzi parameter $\gamma$ and the polymeric parameter $\delta$, i.e., $\epsilon=\gamma \delta \ll 1$. The parameter $a_{0}$ is defined as
\bqn
a_0 = \frac{A_{\rm min}}{8\pi},
\eqn
where $A_{\rm min}$ represents the minimum area gap of LQG. It is interesting to mention that $A_{\rm min}$ is related to the Planck length $l_{\rm Pl}$ through $A_{\min} \simeq 4 \pi \gamma \sqrt{3} l_{\rm Pl}^2$ \citep{Sahu:2015dea}. Thus, $a_0$ is proportional to $l_{\rm Pl}$ and is expected to be negligible. Hence, phenomenologically, the effects of $a_0$ on spacetimes are expected to be very small, and in this paper, we will only focus on the effects on the scale of a few to thousands of Schwarzschild radii of LQG and set $a_0=0$. The gravitational constant in the rotating LQBH ($G_\text{LQG}$) can be related to the Newtonian gravitational constant ($G_{\rm N}$) by 
\bqn
G_\text{LQG}=G_{\rm N}\frac{(1+P)^2}{(1-P)^2}.
\eqn  
Finally, the event horizon for rotating LQBH can be obtained by solving $\Delta(r_{h\pm})=0$, where the "+" denotes the outer horizon and "-" denotes the inner horizon, and the explicit expression of $r_{h\pm}$ can be found in \cite{Liu:2020ola}. Note that, for $P=0$, $r_{h\pm}$ transitions to the widely recognized expression for the Kerr black hole, i.e., $r_{h\pm}=M \pm \sqrt{M^2-a^2}$.

In our numerical simulations, we substitute the Newtonian gravitational constant into our metric (Eq.~(\ref{metric})) to make a fair comparison between LQBH and Kerr BH (i.e., $G=G_{\rm LQG}$). Considering that dimensionless spin $a$ is also proportional to the gravitational constant, which also needs to be scaled by a factor of ${(1+P)^2}/{(1-P)^2}$. We should note that \cite{Afrin:2022ztr} uses the gravitational constant $G_{\rm LQG}$ rather than the Newtonian gravitational constant $G_{\rm N}$ in their comparison between the shadow image data and LQBH with different polymeric functions.
Therefore, the result of our work is expected to be different from \cite{Afrin:2022ztr}.

Finally, the Immirzi parameter $(\gamma)$ has a lot of choices from different considerations \citep[see][]{BenAchour:2014qca, Frodden:2012dq, Achour:2014eqa,
Han:2014xna,Carlip:2014bfa, Taveras:2008yf}. It has been shown that its value can even be a complex number \citep{Frodden:2012dq, BenAchour:2014qca,  Carlip:2014bfa,Meissner:2004ju}, or considered as a scalar field in which the value would be fixed by the dynamics \citep{Taveras:2008yf}. 
We adopt the commonly used value $\gamma = 0.2375$  from the black hole entropy calculation \citep{Meissner:2004ju}. 


\begin{figure}
    \centering
    \includegraphics[width=0.8\linewidth]{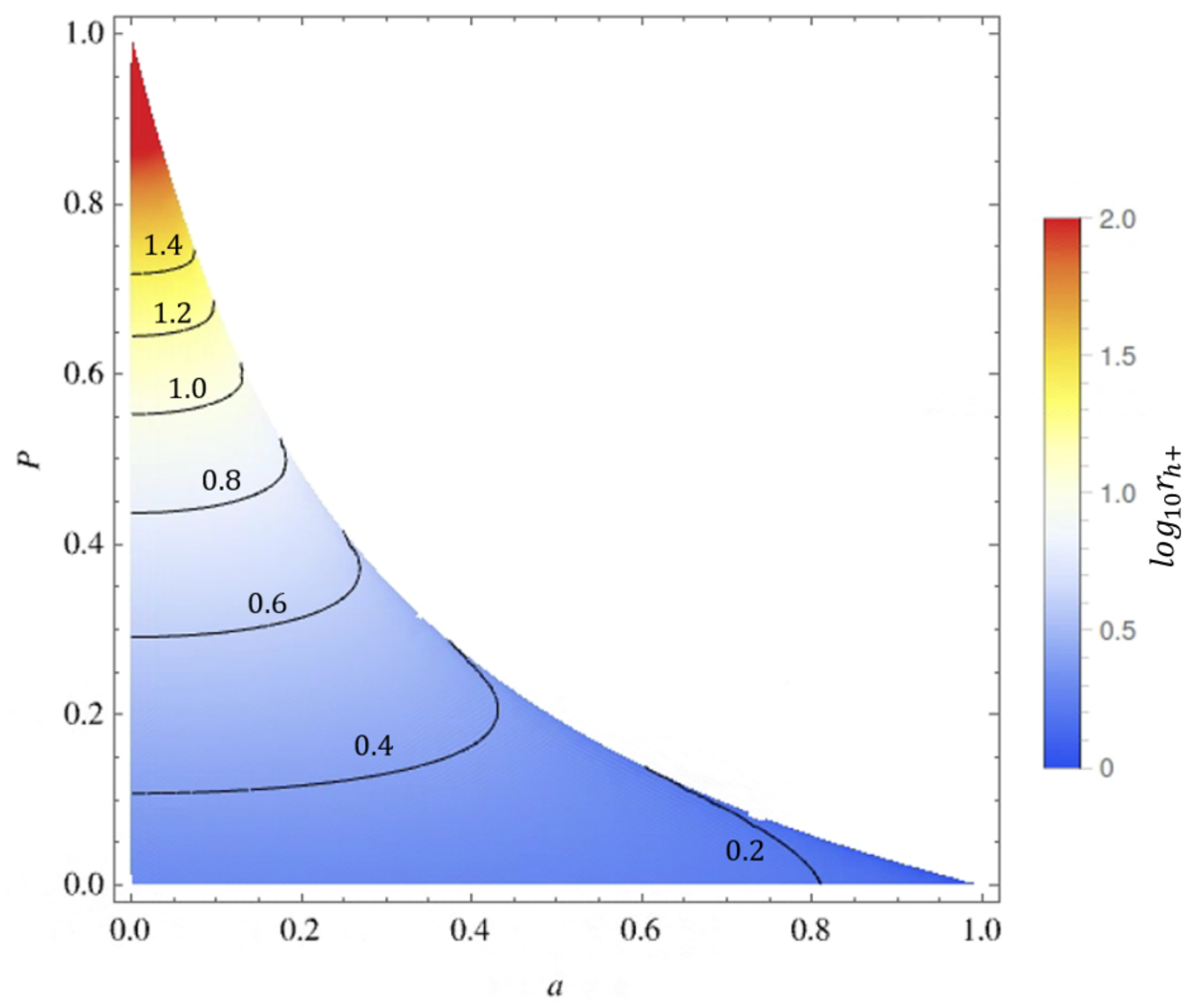}
    \caption{Logarithmic distribution of the outer horizon $r_{h+}$ {in the unit of $G_{\rm N} M/c^2$} on polymeric function $P$ and spin parameter $a$ plane. 
    }  
    \label{horizon_contour}
\end{figure}

Figure~\ref{horizon_contour} shows the radius of the outer horizon on the $(P, a)$ plane. This helps us see how the outer horizon is connected to the polymeric function $P$ and the spin $a$. The color represents the radius of the outer horizon. With fixed $a$, the radius of the outer horizon increases initially with $P$. However, with a further increase in $P$, the radius of the outer horizon saturates and then decreases with $P$. Thus, the value of the outer horizon shows degenerate values in a certain range of $P$ for any given spin parameter ($a$).

\section{Semi-analytical simulation of shadow images of LQBH}

As the geodesic image of photons in BH spacetime appears to a distant observer, the BH shadow reflects the BH spacetime properties. By comparing with EHT observations \citep{EventHorizonTelescope:2022wkp, EventHorizonTelescope:2022apq, EventHorizonTelescope:2022wok, EventHorizonTelescope:2022exc, EventHorizonTelescope:2022urf, EventHorizonTelescope:2022xqj}, we can constrain the value of polymeric function $P$ in LQG. Although analytical calculations \citep{Afrin:2022ztr, Liu:2020ola} can determine the shadow silhouette cast, in reality, the accretion disk produces extended emission around the photon ring, which significantly alters the shadow morphology. The impact of plasma and metric on the BH images was investigated earlier \citep{Ozel2022, Younsi2023}, which shows the size of the luminous ring is greatly influenced by both of them. Therefore, to make it more realistic, we implement semi-analytical simulations to investigate the radiative properties of the accretion flow around an LQBH and use EHT results to constrain the loop quantum effect. 

\subsection{Accretion flows and camera setup}


In this subsection, we describe the accretion flow model and the camera setup that we use to generate the shadow images of LQBH. The shadow images in this work are generated from a semi-analytical optically thin Radiatively Inefficient Accretion Flow (RIAF) \citep{Yuan:2003dc,
Pu:2018ute} in the LQBH spacetime. The simulations are implemented with a polarized General Relativistic Ray Tracing (GRRT) technique using the code {\tt ipole}\footnote{\url{https://github.com/AFD-Illinois/ipole}} \citep{Moscibrodzka:2017lcu, Noble:2007zx}. 
To generate BH shadow images, the calculation sets a camera $1\,000\,\rm r_{\rm g}$ away from the BH with a field of view (FoV) of $200\times200\,\rm \mu as$. By calculating the geodesic motion of the photons, we simulate the imagining process in the spacetime of LQG and obtain the shadow image. The resolution of the images is $1\,000\times1\,000$ pixels. The observing frequency is $230\,\rm GHz$. For the simulations in sections 3, 4.1, and 4.2, we use Sgr\,A$^*$ as the target. The black hole mass is set to be $4\times10^6\,\rm M_{\odot}$ with a distance of $8\,\rm kpc$ which comes from \cite{EventHorizonTelescope:2022wkp}. Since we only focus on the $230\,\rm GHz$ emission in this work, non-thermal emission is less dominant. We consider a thermal electron distribution function for the calculation of the synchrotron emissivity.

Multiple parameters about the accretion flow are related to the nature of the accretion flow, which include electron number density $n_{\rm e}$, electron temperature distribution $T_{\rm e}$, and disk thickness $H$. The dynamics of the accretion flows also influence the polarization of the shadow images. 
Based on current understanding, magnetically arrested disk (MAD) is the best-bet model from the EHT observations \cite{EventHorizonTelescope:2022urf}, accordingly, we consider a MAD mimicking disk configuration following \cite{Chen:2021lvo}. 
For simplicity, we consider the plasma inside of the innermost stable circular orbit $R_{\rm ISCO}$ to be in a free fall, while outside of $R_{\rm ISCO}$ rotates with sub-Keplerian velocity. The radius of ISCO is numerically solved with the equation of motion in Appendix~\ref{sec:geodesic} and input into {\tt ipole} manually for each simulation.
Since the angular frequency caused by free fall is $\omega_{\rm FF} = -g_{t\phi}/g_{\phi\phi}$, Keplerian angular frequency $\omega_{\rm K}=\sqrt{-g_{tt,r}/g_{\phi\phi,r}}$. The sub-Keplerian motion for the accretion disk in our model is defined as
 \begin{equation}
     \omega_{\rm SubK}=\omega_{\rm K} + (1 - K)(\omega_{\rm FF} - \omega_{\rm K}), \label{Eq:subKepler}
 \end{equation}
where $K$ is the Keplerian factor. Since the free fall radial velocity is $u^r_{\rm FF} = \sqrt{|-(1 + g_{tt})/g_{rr}|}$. For the sub-Keplerian setup, inflow $u^r$ satisfies $u^r_{\rm SubK}=(1-K)\,u^r_{\rm FF}$. 
 
 In our semi-analytical GRRT calculation, we adopt the default setting of electron number density ($n_{\rm e}$ unit $3\times10^7\,\rm cm^{-3}$) and temperature ($T_{\rm e}$ unit is $3\times10^{11}\,\rm K$) for Sgr\,A$^*$, which generates a total flux ranging from $1-3\,\rm Jy$ depending on parameters like the torus thickness $H$, line of sight inclination angle $i$, etc. For our fiducial accretion disk setup, which uses the MAD-like setup in \cite{Chen:2021lvo}, we fix $K=0.5$, disk thickness $H=0.3$. 

 \subsection{Enlarged ring from LQBH}
 
In the previous section, we discussed the radius distribution of the outer horizon (see Fig.~\ref{horizon_contour}). The polymeric function $P$ affects the radius of the horizon $r_{\rm H}$ differently depending on the dimensionless spin $a$. In general, the outer horizon expands with $a\lesssim 0.3$ and shrinks with $a\gtrsim0.8$. Similar expansion for the ring diameter happens in the ray-traced shadow images and the photon sphere. 
 
\begin{figure}
     \centering
     \includegraphics[height=0.35\linewidth]{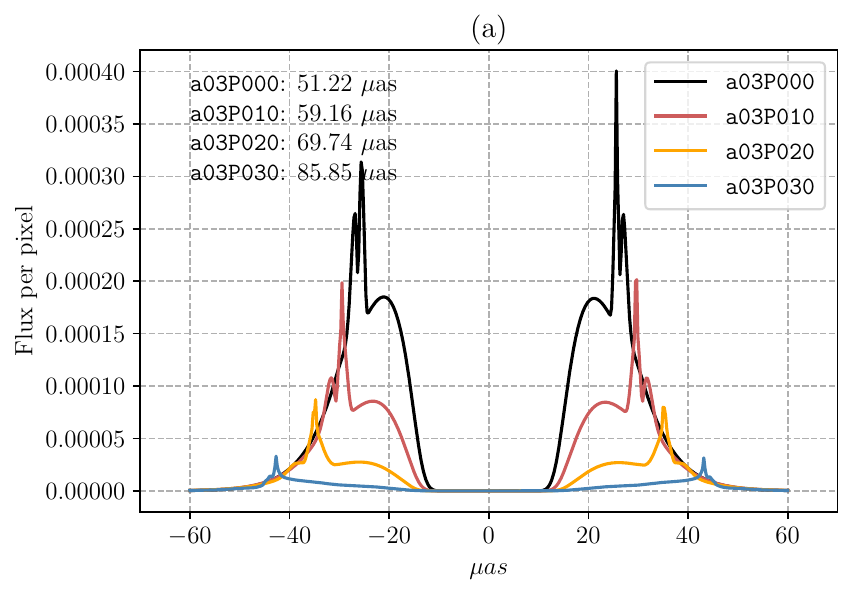}
      \includegraphics[height=0.35\linewidth]{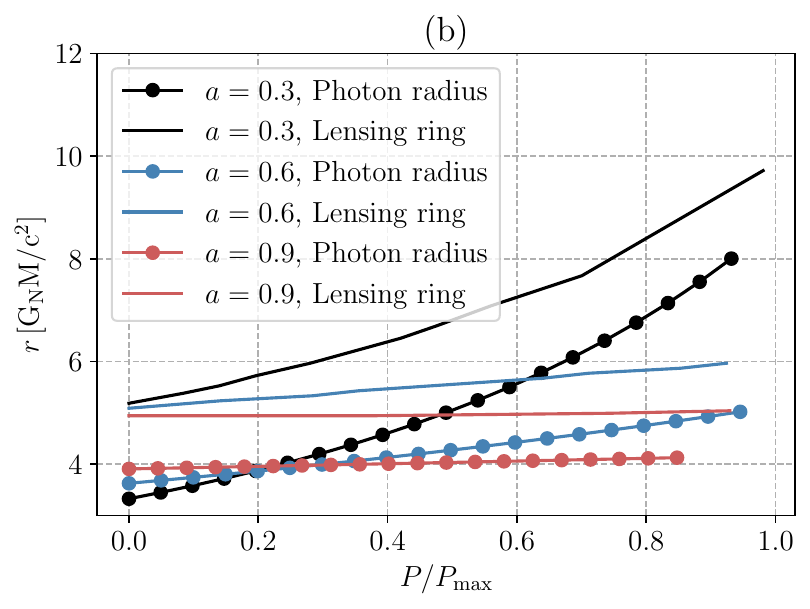}
     \caption{Panel (a): Cross-cut profiles of the $a=0.3$ cases with $P=0$ {(black)}, $0.1$ {(red)}, $0.2$ {(orange)}, and $0.3$ {(blue)}. Panel (b): The radius of photon sphere and ring diameter evolution with increasing polymeric function $P$. The black, blue, and red solid lines represent the BH with $a=0.3$, 0.6, and 0.9, respectively. The dotted solid lines are the radii of the lensing rings.
     }
     \label{fig:dsh_expansion}
 \end{figure}

We use an inclination angle of $1^\circ$ to get closely symmetric shadow images, where the influence by the BH spin is minimal.
To quantitatively determine the shadow size, following \cite{Bronzwaer:2020kle}, we calculate the cross-cut profiles of the intensity maps \footnote{Defining the intensity map as $I(X, Y)$, the cross-cut profile is $I(X,0)$.}. The distance between the two peaks (lensing ring) in the cross-cut profile is the ring diameter $d_{\rm ring}$. We present the cross-cut profile of the $a=0.3$ with different $P$ cases in Fig.~\ref{fig:dsh_expansion}a. We observe a significantly larger ring diameter and less brightness asymmetry in the case with $P=0.3$. 
We measure the ring diameter with a series of GRRT images using the same method, with different spin values, $a=0.3$, $0.6$, and $0.9$, and with increasing the value of $P$. The radius of the unstable photon orbit $R_{\rm photon}$ and the radius of the photon ring are presented in Fig.~\ref{fig:dsh_expansion}b. Due to the different maximum polymetric function $P_{\rm max}$ for different BH spins, we scale the $P$ in the form of $P/P_{\rm max}$ in Fig.~\ref{fig:dsh_expansion}b to plot the lines in one figure. As \cite{Younsi2023,Ozel2022} suggested, the lensing ring in the shadow image and the photon sphere are not exactly the same, the scaling relation between them has been studied in \cite{Akiyama:2019eap, EventHorizonTelescope:2022xqj, Younsi2023}. Fig.~\ref{fig:dsh_expansion}b shows both of them, they increase monotonously with $P$, which is different from the trend of the outer horizon. 

 \subsection{Polarimetry from LQBH}
 
\begin{figure}
    \centering
    \includegraphics[height=0.33\linewidth]{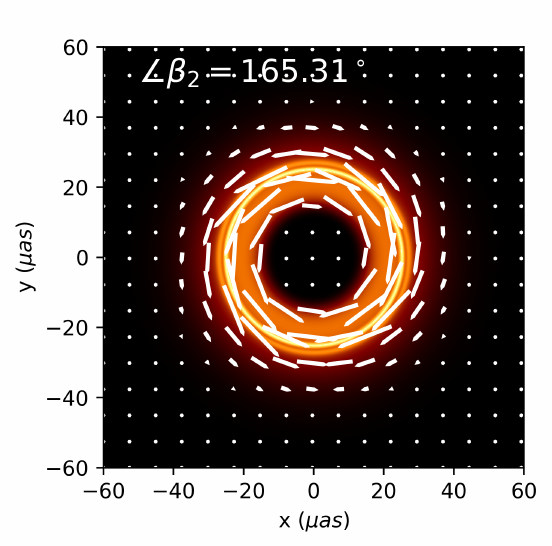}
    \includegraphics[height=0.33\linewidth]{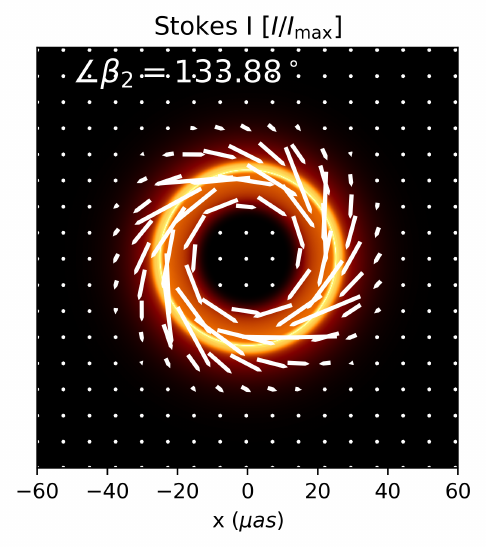}
    \includegraphics[height=0.33\linewidth]{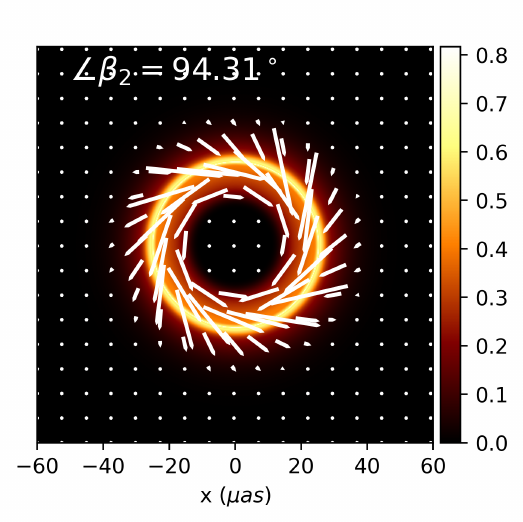}
    \caption{Intensity map ($I/I_{\rm max}$) along with polarimetric pattern of free fall ({\it left}), sub-Keplerian ({\it middle}), and Keplerian disk ({\it right}) from a Kerr BH with $a=0.6$.}
    \label{fig:polarization}
\end{figure}

Polarimetric shadow images of M87\,$^*$ were released in \cite{EventHorizonTelescope:2021bee}, which provides information on the magnetic field configuration close to M\,87. \citep{EventHorizonTelescope:2021srq}. Although the polarimetric information of Sgr\,A$^*$ has not been released yet, it is still worth comparing the polarimetric image between LQBH and Kerr BH with Sgr\,A$^*$ as a target. We apply our fiducial model for the accretion flows and an inclination angle of $1^\circ$ for GRRT simulations. We study the polarization properties within the framework of LQG. 

In this paper, we use the International Astronomical Union (IAU) definition of Stokes parameters $(\mathcal{I, Q, U ,V})$ \citep{1996A&AS..117..161H, Smirnov:2011vp}.
 To do a quantitative analysis of the polarimetric pattern. We adopt the definition of complex linear polarization in \cite{EventHorizonTelescope:2021bee, EventHorizonTelescope:2021srq}:
 \begin{equation}
     \mathcal{P} = \mathcal{Q} + i\mathcal{U}.
 \end{equation}
The electric-vector position angle (EVPA) is defined as 
\begin{equation}
    {\rm EVPA} \equiv \frac{1}{2}{\rm arg(\mathcal{P})}.
\end{equation}
 To quantitatively measure the EVPA pattern across the image, we adopt the decomposition of the complex linear polarization $\mathcal{P}$ into azimuthal modes with complex coefficients $\beta_{\rm m}$ \citep{EventHorizonTelescope:2021srq, Palumbo:2020flt, Chael:2023pwp}. The definition of $\beta_{\rm m}$ is
 \begin{equation}
     \beta_{\rm m}=\frac{1}{I_{\rm ann}}\int_{r_{\rm min}}^{r_{\rm max}}\int_0^{2\pi} 
     \mathcal{P}(r,\phi)\exp{(-im\phi)}rd\phi dr,
 \end{equation}
where $r_{\rm max}$ and $r_{\rm min}$ are the boundaries of the ring region. We follow the procedure of ring extraction in \cite{EventHorizonTelescope:2022exc, Palumbo:2020flt}. First, we identify the ring center. We use 360 azimuthal slices to identify the ring center as the point that has the same distance from the peak emission in each slice. Then, we calculate the ring width by taking the mean of the full width along each azimuthal angle from the ring center. Note that the full width is calculated if the intesity $I/I_{\rm max}>0.05$. We include most of the luminous region for the ring to smooth the influence from the accretion flows and focus on the effect of LQG. 
Since both Sgr\,A$^*$ and M87\,$^*$ have small inclination angles, the rotationally symmetric mode $\beta_2$ is the dominant mode in our case. The complex phase of $\beta_2$ mode $\angle \beta_2$ quantitatively describes the global EVPA pattern in the shadow images.  
\begin{figure}
     \centering
     \includegraphics[width=0.47\linewidth]{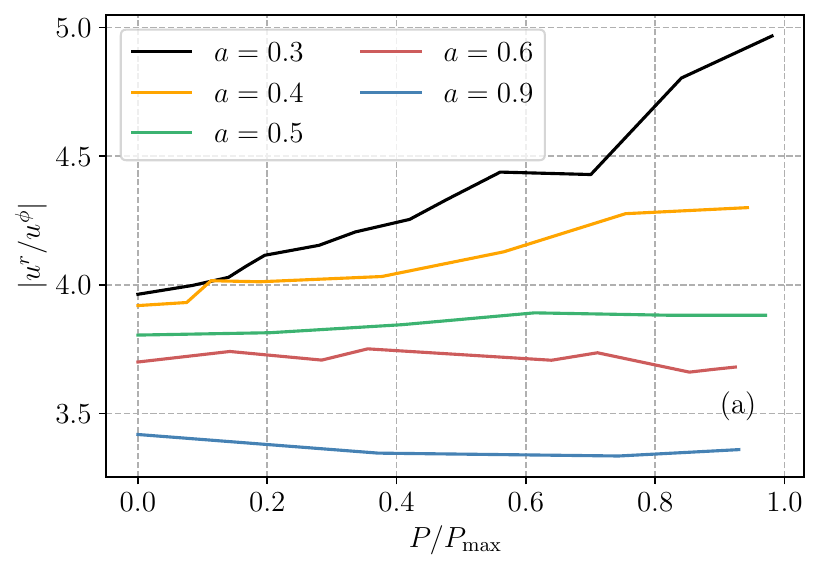}
     \includegraphics[width=0.47\linewidth]{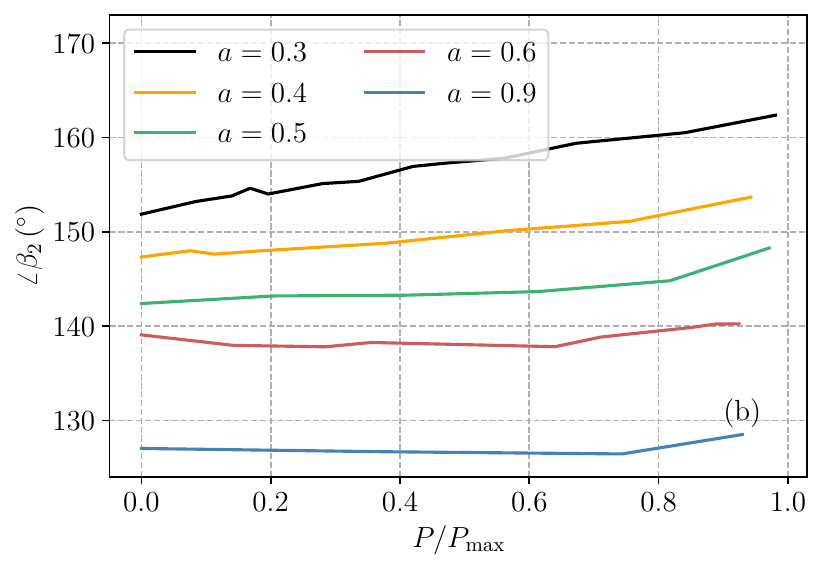}
     \caption{Panel (a) represents the ratio between radial and circular motion at the position of the lensing ring. The black line represents the case of LQBH with $a=0.3$; the red line represents the case $a=0.6$; the blue line is the cases with $a=0.9$. Panel (b) represents the complex phase of $\beta_2$ mode $\angle \beta_2$ with different $P$. The black, red, and blue lines represent the cases with $a=0.3$, 0.6, and 0.9, respectively.}
     \label{fig:ur_uph}
 \end{figure}
The polarimetric pattern is related to the motion of the accretion flows and the configuration of the magnetic field. Using a vertical magnetic field configuration, we present the shadow images $(I/I_{\rm max})$ and polarization pattern of free fall, sub-Keplerian, and Keplerian accretion disks of a Kerr BH with $a=0.6$ in Fig.~\ref{fig:polarization}. This figure shows how the dynamics of the accretion flows influence the polarization angle. A similar result is also pointed out in \cite{EventHorizonTelescope:2021srq}. The three types of motion correspond to the Keplerian factors $K=0$, $0.5$, and $1$ in Eq.~\ref{Eq:subKepler}, respectively, which leads to different ratios of the radial and angular motions ($|u^r/u^\phi|$). It is expressed as
 \begin{equation}
     |u^r/u^\phi| = \frac{(1-K)u^r_{\rm FF}}{\left[\omega_{\rm K}+(1+K)(\omega_{\rm FF}-\omega_{\rm K})\right]u^t},
 \end{equation}
where $u^t=\sqrt{|-(1+ {(u^r)}^2 g^{rr})/\mathcal{C}|}$ and $\mathcal{C}$ is the carter constant. 
Considering most of the emission comes from the lensing ring, the calculation of $\beta_2$ is strongly related to this region because $\beta_2$ is scaled by $\mathcal{I}$. 
The property at the lensing ring is representative of the whole luminous ring. Using this radius, the values of $|u^r/u^\phi|$ of the 3 cases are 36.63, 3.70, and 0, respectively. The corresponding $\angle \beta_2$ are $165.31^\circ$ ,$133.88^\circ$, and $94.31^\circ$. 

Since polarization is also related to spacetime, we implemented another case of LQBH, which shares the same spin parameter $a=0.6$, but uses an extreme polymetric function $P=0.16$. The accretion flows are set to be free-falling. In this case, the value of $\angle \beta_2$ is $165.10^\circ$. We obtain a similar value to that in the Kerr BH case. It indicates that it does not have any significant influence from the different spacetimes. Therefore, different accretion flow structures have a more significant effect on the polarimetric pattern. It shows that radial motion leads to a higher $\angle \beta_2$ and a circular polarimetric pattern. The increment of angular motion in the $\phi$ direction increases the radial component of the polarimetric pattern, which decreases $\angle \beta_2$.
 
The impact of LQG on the value of $\angle\beta_2$ by using different spins with $a=0.3$, $0.6$, and $0.9$ is presented in Fig.~\ref{fig:ur_uph}b. As it shows, with $a\lesssim 0.5$, the value of $\angle\beta_2$ increases monotonically with the increment of $P$. It implies that the photons come from a region with less circular motion. 
For sub-Keplerian motion, the ratio between radial and circular motion $|u^r/u^\phi|$ is usually increases with radius. At the position of the lensing ring, we plot the dependence of the $|u^r/u^\phi|$ ratio on $P$ in Fig.~\ref{fig:ur_uph}a. In the $a=0.3$ case (black line), as $P$ increases, the contribution of radial motion increases monotonically. Recalling the case of free-fall in Fig.~\ref{fig:polarization}, it has a relatively high $\angle \beta_2$. This explains the increasing trend of $\angle\beta_2$ in the $a=0.3$ cases seen in Fig.~\ref{fig:ur_uph}a. However, as $a$ increases to $\sim 0.9$, $|u^r/u^\phi|$ starts to decrease with increasing $P$. It implies a higher circular motion for the LQBHs compared with Kerr BH, which leads to a slightly lower $\angle\beta_2$ in Fig.~\ref{fig:ur_uph}a.

In conclusion, the value of $\angle \beta_2$ is very useful to determine the spin parameter of BHs. However, the difference between $\angle\beta_2$ may be minor among different $P$ especially when $a \gtrsim 0.5$.

\section{Constraints of Shadow Size from EHT Observation}\label{sec:EHT_constraint}

From the observation of Sgr~A$^*$, EHT collaboration made multiple constraints on the properties of the SMBH at the center of our galaxy \citep{EventHorizonTelescope:2022wkp,
EventHorizonTelescope:2022apq,
EventHorizonTelescope:2022wok,
EventHorizonTelescope:2022exc,
EventHorizonTelescope:2022urf,
EventHorizonTelescope:2022xqj}. One of the strongest constraints comes from the measurement of the angular ring diameter $d_{\rm ring}$, which is $51.8\pm2.3\,\rm \mu as$ \citep{EventHorizonTelescope:2022wkp}. In the case of $P\ne0$, LQBH possesses different sizes horizon as well as the photon ring for the same spin parameter (see Sec.~\ref{sec:LQBH}). Direct comparison between the observed $d_{\rm ring}$ and theoretical result puts an upper limit on the value of $P$. Previous analytical work \citep{Afrin:2022ztr} concluded $P\lesssim 0.0423$. However, as we point out in the previous section, the previous study lacks the consideration of the modification of the gravitational constant. In this section, we include the effect of the gravitational constant for consistency. Meanwhile, we consider a more realistic regime using {\tt ipole} simulation with our fiducial model which is the same as the MAD torus in \cite{Chen:2021lvo}. We performed a parameter survey to determine the LQBH parameters (the details of the parameters of each run see Appendix~\ref{Sec:survey}). 

\subsection{Parameter survey}
\begin{figure*}
    \centering
    \includegraphics[height=0.48\linewidth]{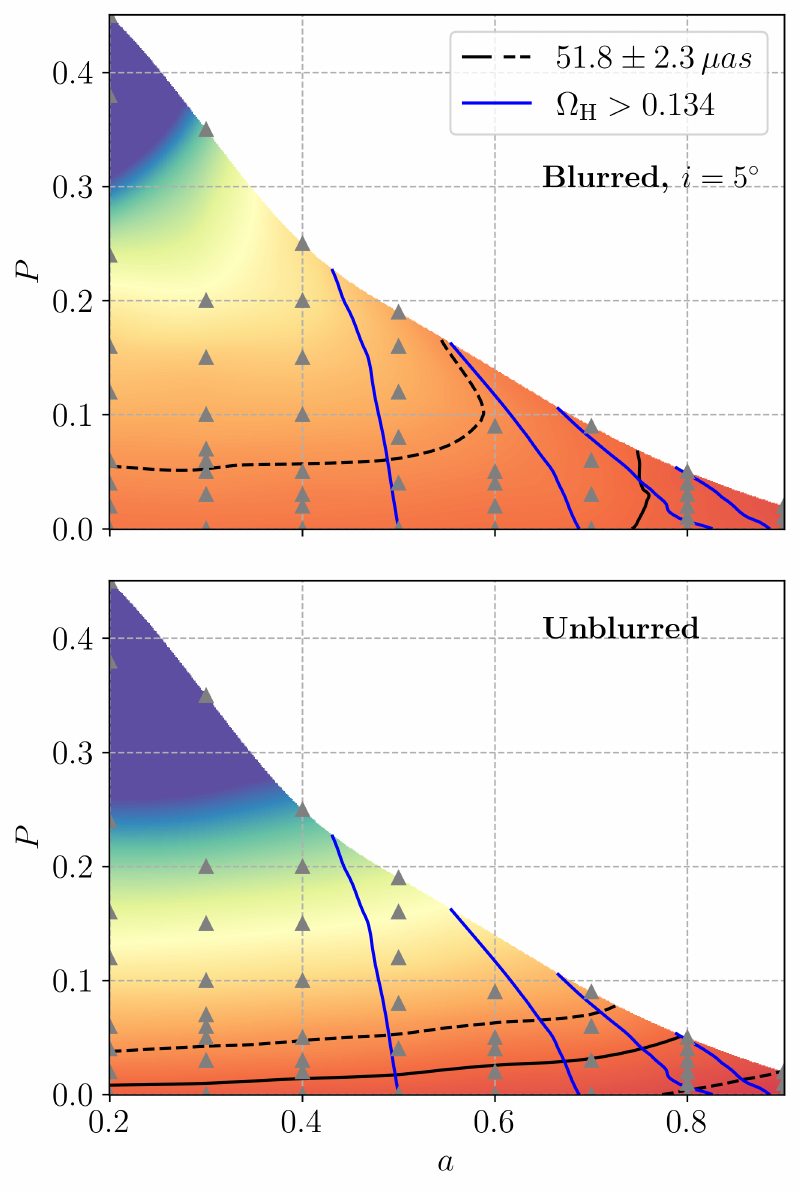}
    \includegraphics[height=0.48\linewidth]{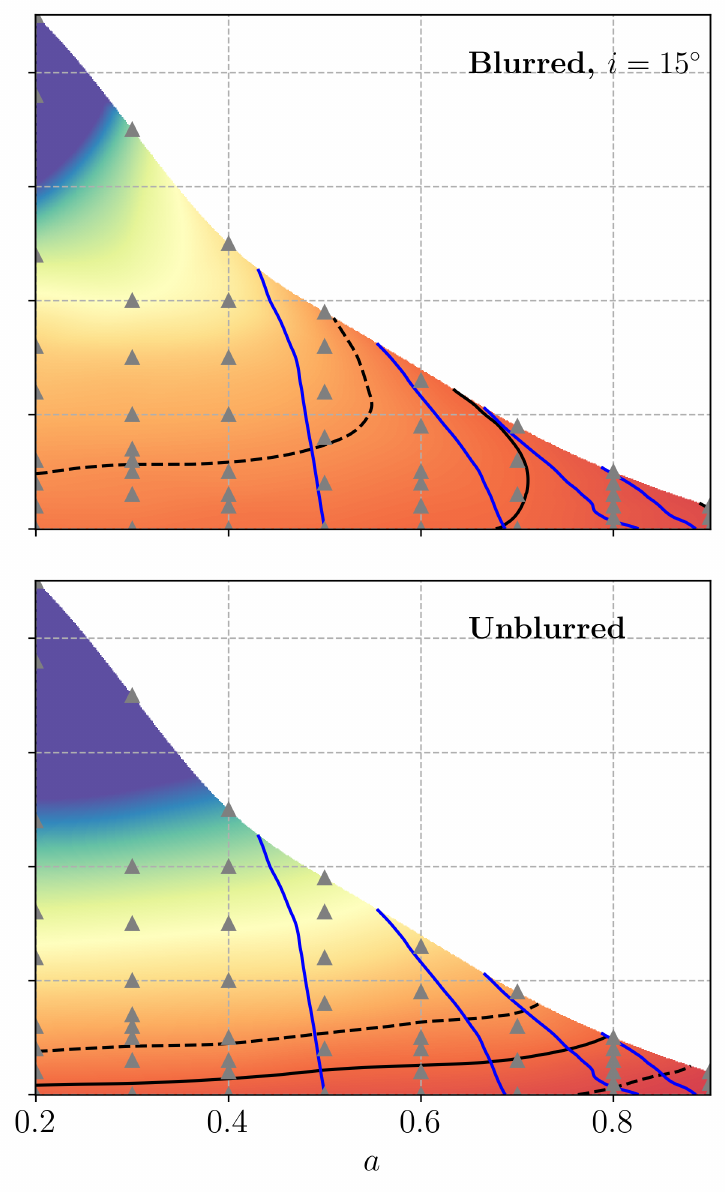}
    \includegraphics[height=0.48\linewidth]{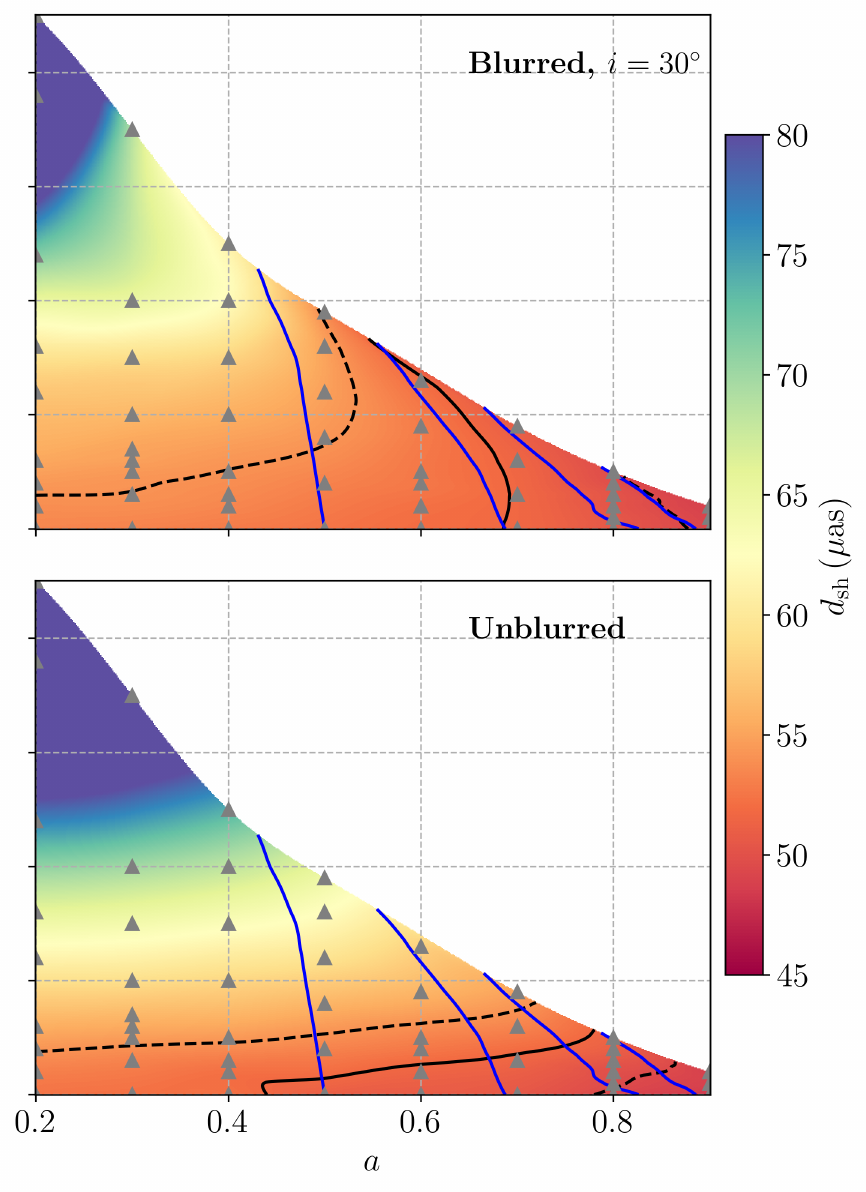}
    
    \caption{The shadow size from LQBHs with different dimensionless spin $a$ and polymeric function $P$. The first row shows the ring diameter from blurred shadow images, while the second row is from the original shadow images. The {\it left}, {\it middle}, and {\it right}  columns are for inclination angle of $5^{\circ}, 15^{\circ},$ and $30^{\circ}$, respectively. See text for more details.}     \label{fig:dsh}
\end{figure*}

\begin{figure*}
    \centering
    \includegraphics[width=\linewidth]{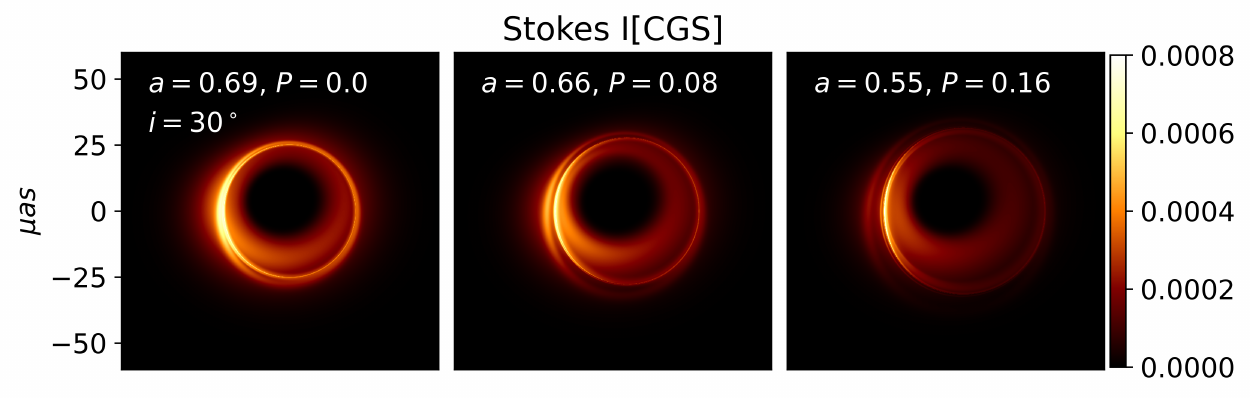}
    \includegraphics[width=\linewidth]{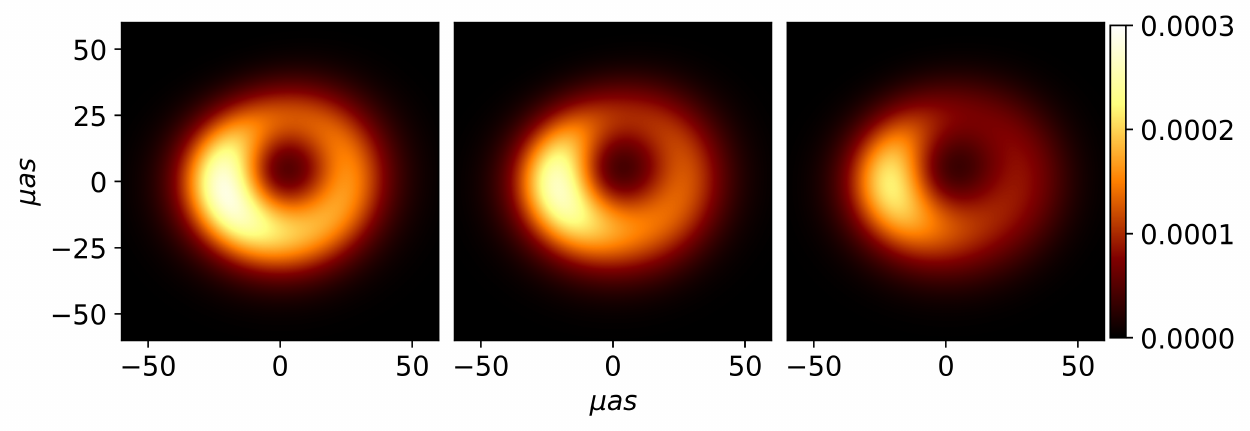}
    \caption{Shadow images for different BH parameters of {\it left} ($a=0.69$, $P=0.0$), {\it middle} ($a=0.66$, $P=0.08$), and {\it right} ($a=0.55$, $P=0.16$) with $i=30^\circ$ shown in the {\it upper} panels . The {\it lower} panels show the blurred shadow images for the same sets of parameters.}     \label{fig:obs_par}
\end{figure*}

The polymeric function $P$ directly impacts the sizes of the shadow images. The higher value of $P$ generally generates a larger photon ring and shadow size. Using the same method in the previous section, we measure the ring diameter $d_{\rm ring}$ from the GRRT result. 
However, as discussed in \cite{EventHorizonTelescope:2022xqj}, the observed shadow diameter $\hat{d}_{\rm sh}$ is different with real shadow diameter $d_{\rm ring}$ by a calibration factor $\alpha_{\rm c}$, i.e., $\hat{d}_{\rm sh}=\alpha_{\rm c} d_{\rm sh}$ (for detail see \cite{EventHorizonTelescope:2022xqj}). A similar deviation also happens to the measurement of ring diameter. The limited resolution strongly influences the measurement of the ring (see Fig.~\ref{fig:dsh}).
A more self-consistent way is blurring the theoretical result to the resolution of EHT and comparing them with the observation. Both original shadow images and blurred ones are obtained in this work. The blurred images are convolved by a Gaussian beam of FWHM with $20\,\rm \mu as$, which is the typical resolution of EHT \citep{EventHorizonTelescope:2022xqj, Chael:2018oym}.

We interpolate the ring diameters produced from different BH parameters and show them in Fig.~\ref{fig:dsh}. The grey triangles represent the GRRT simulations we implemented. The red solid and dashed lines show the contour for EHT observation and its $1\sigma$ uncertainty contour, respectively. The blue contours correspond to the region with angular momentum that is larger than that of a Kerr BH with $a=0.5$. From left to right, the three columns of Fig.~\ref{fig:dsh} represent the result from line-of-sight inclination angles of $5^\circ$, $15^\circ$, and $30^\circ$. 
The panels in the upper row correspond to the blurred images, while those in the lower row correspond to the unblurred images.
The large uncertainty from EHT observations covers a large region of parameter space.
By comparing different panels, we observe that the allowed region moves to a lower spin parameter as the inclination angle increases from $5^\circ$ to $30^\circ$. However, the differences between the allowed ranges of spin parameters for the different inclination angles are insignificant. 
The difference between the value of BH spins $a$ at the EHT constrain, ${d}_{\rm ring}=51.8\,\rm \mu as$ in inclination angle $i=5^\circ$ and that in $i=30^\circ$ is $\sim 0.1$.
Considering the different results from unblurred and blurred shadow images, this result is influenced by the limited resolution of EHT observation. 

The parameter ranges obtained from the lower row of Fig.~\ref{fig:dsh} differ from the blurred ones in the upper row. 
The problem of the unblurred cases is a direct comparison between theoretical results and observation results, which may be unreliable due to the limited observational resolution. Accordingly, from the unblurred results, we conclude the maximum value for $P$ is $\sim 0.1$ within $1\sigma$ uncertainty. 
The blurred results cover a wide range of $P$ values. Therefore, it is hard to determine the maximum allowed value of $P$ only from the shadow size at the current EHT observations. Nonetheless, from the blurred images, we note that the allowed value of $P$ could be of the order of $P\lesssim0.2$ for Sgr~A$^*$.

The value of $P$ for Sgr~A$^*$ will be constrained better with the help of the angular momentum of the black hole. The angular frequency at the event horizon is expressed as:
\begin{equation}
    \Omega_{\rm H} = \left(-\frac{g_{t\phi}^{\rm BL}}{g_{\phi\phi}^{\rm BL}}\right)_{r_{\rm H}},
\end{equation}
where $r_{\rm H}=r_{h+}$ is the radius of the outer horizon. 
Previously, it has been shown that a relatively high value of black hole spin is favored for Sgr~A$^*$ \citep{EventHorizonTelescope:2022urf}.
If we put a lower limit for the BH spin parameter $a \geq 0.5$, the minimum value for $\Omega_{\rm H}$ is 0.134 for a Kerr BH. Therefore, we consider the lower limit of the angular frequency of the Sgr~A$^*$ to be $\Omega_{\rm H}\geq 0.134$. In Fig.~\ref{fig:dsh}, the crossing region of the blue and red contours gives allowed parameter ranges for $P$. 
The diameters of the blurred shadow images are largely related to the size of the outer horizon. For $a>0.5$, the change in the outer horizon is minimal with the increases of $P$ (see Fig.~\ref{horizon_contour}). Therefore, the $d_{\rm ring}=48.7\,\mu as$ contour is roughly vertical. Although the radius of the photon ring is expected to be large for different $P$, the limited resolution of EHT is not enough to resolve the differences between them. With the limited resolution of EHT, we can conclude that Sgr\,A$^*$ can have values of $P\lesssim 0.2$ considering $a\gtrsim 0.5$.

Besides setting thickness $H=0.3$, we also test two different values, $H=0.1$ and $0.05$ for the parameter survey. Both blurred and unblurred result shows similar constraint to the $a$ and $P$ with a negligible difference ($\sim 0.01$). Therefore our result strongly suggest the value of polymetric function for Sgr\,A$^*$ to be $P\lesssim 0.2$.

\subsection{Degeneracy of blurred shadow}

To understand more detail of the shadow image of Sgr~A$^*$ with LQBH. We select 3 points on the $d_{\rm ring}=51.8\,\mu as$ contour of the blurred cases with an inclination angle of $30^\circ$, which correspond to similar ring diameters in the resolution of EHT (upper right panel of Fig.~\ref{fig:dsh}). Their parameters are $(a,P)=(0.69, 0.00)$, $(0.66, 0.08)$, and $(0.55, 0.16)$, where the first number in the brackets is the value of BH spin while the second one is the value of the polymeric function. The GRRT results of these parameters are presented in Fig.~\ref{fig:obs_par}. The upper panels are the unblurred shadow images, while the lower ones are blurred results. As the $P$ increases, we can clearly see a larger photon ring. The diameters of the lensing ring are $51.45, 54,25$ and $60.66\,\rm \mu as$.  However, it is not clearly reflected in the corresponding blurred images. The radii of the outer horizons of the 3 black holes are $1.73\,\rm r_{\rm g}$, $1.79\,\rm r_{\rm g}$, and $1.87\,\rm r_{\rm g}$, corresponding to ring diameters of $51.45, 51.05$, $51.05\,\rm \mu as$. To distinguish the degeneracy of spin and polymeric function, observations in higher resolution are required, which may be obtained from future EHT and ngEHT observations \citep{Tiede:2022grp, Johnson:2023ynn}. From Fig.~\ref{fig:obs_par}, we also see that the BH images in the right column are slightly dimmer than those of the other two columns. 
Comparing the blurred images, we find that the brightest pixel on the right panel ($a=0.55$, $P=0.16$, LQBH case) is about $77\%$ of the left panel ($a=0.69$, $P=0$, Kerr BH case). 

The value of $\angle\beta_2$ is another channel to distinguish the Black Holes. The $\angle\beta_2$ that calculated from unblurred GRRT result are $125^\circ$, $123^\circ$, and $124^\circ$, and from the blurred result, $\angle\beta_2$ are $129^\circ$, $126^\circ$, and $123^\circ$. The difference of $\angle\beta_2$ is too small to be distinguished in current polarimetric observation \cite{EventHorizonTelescope:2021srq}. Therefore the measurement of $\beta_2$ in the case of Sgr\,A$^*$ may not be enough to determine the $a$ and $P$ parameters. Observations with higher angular resolutions of Sgr~A$^*$ will be important to find a more consistent upper limit of LQG parameter $P$ in future surveys (e.g., ngEHT).

\subsection{EHT constraint on the polymetric function of M\,87$^*$}
\begin{figure}
    \centering
    \includegraphics[width=\linewidth]{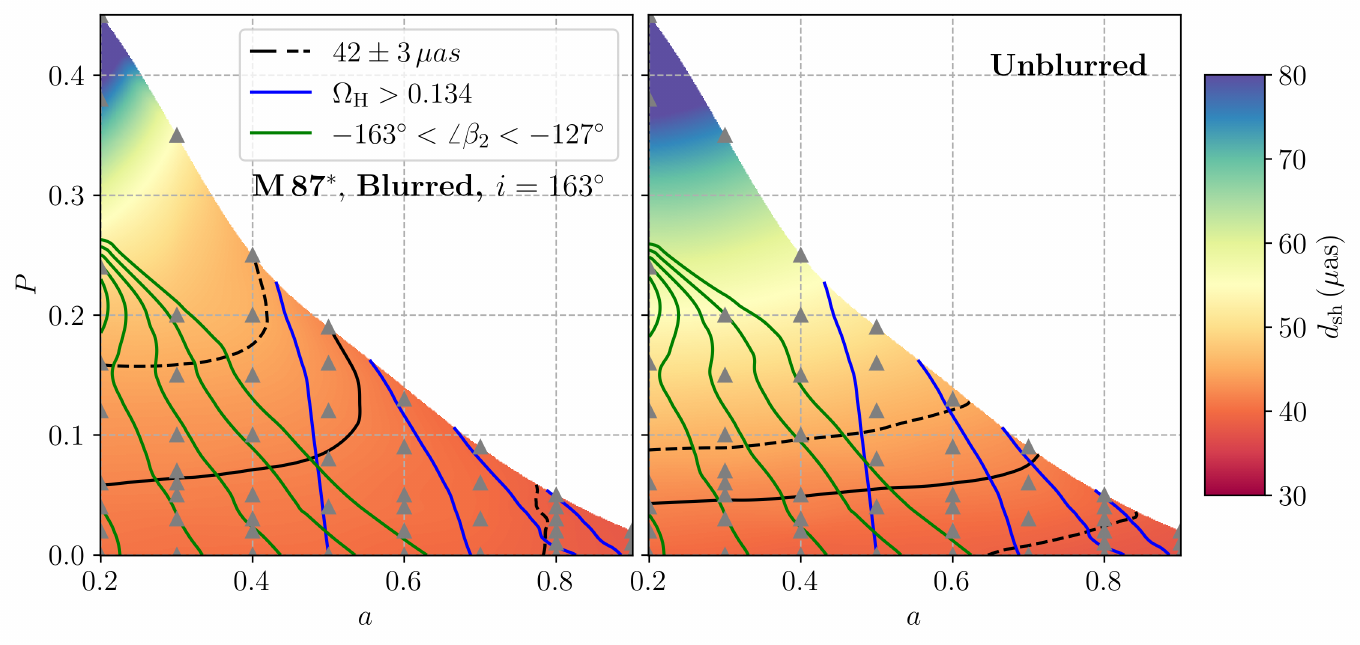}
    \caption{The same with Fig.~\ref{fig:dsh} but target is M\,87$^*$. Green contours cover the region $-163^\circ < \angle\beta_2<-127^\circ$. See text for more details.}
    \label{fig:M87}
\end{figure}

M\,87$^*$ is another horizon-scale observed source by EHT \cite{EventHorizonTelescope:2019dse, Akiyama:2019brx, Akiyama:2019eap, Akiyama:2019bqs, Akiyama:2019fyp, Akiyama:2019sww}. EHT observation shows the diameter of the ring of the M\,87$^*$ is $42\pm3\,\mu as$, with a distance of $D=16.8\pm0.8\,\rm Mpc$ and a mass of $M_{\bullet}=(6.5\pm0.7)\times10^9\,\rm M_{\odot}$. 
We can make a direct comparison with observations by using these parameters in our semi-analytical GRRT simulations and setting them up in the same way as in \cite{Chen:2021lvo}: $n_{\rm e} = 1.25\times 10^5\,\rm cm^{-3}$. Further, we can constrain the polymetric function $P$ for M\,87$^*$. 
Thanks to the strong and highly collimated jet in M\,87, its inclination angle is determined as $i \sim 163\pm 2^\circ$ \cite{Walker2018}. We also adopt this value for the GRRT calculation. The mass of M\,87$^*$ $M_{\bullet}$ is $6.5 \times 10^9\,\rm M_\odot$, and the distance is set to be $16.7\,\rm Mpc$ \cite{Akiyama:2019fyp}. 
Although spin $a=0.9 \pm 0.05$ is suggested by \cite{Tamburini:2019vrf} with a $\sim 95\%$ confidence level, the value of $a$ is still under debate and not fully determined yet. Multiple values have been suggested, ranging from 0.5 to $\lesssim1$ \cite{Fromm:2021mqd, Cruz-Osorio:2021cob, Dokuchaev:2019bbf}. 
We assume a minimum spin for M\,87$^*$ that $a \gtrsim 0.5$. Another constraint comes from the measurement of $\angle \beta_2$. EHT polarimetric observations of M\,87$^*$ indicate a vertical magnetic field configuration is more favored to explain the EVPA pattern \citep{EventHorizonTelescope:2021srq}, and the measured absolute value of $-163^\circ < \angle\beta_2<-127^\circ$. 
So, in Fig.~\ref{fig:M87}, we directly compare our simulation of LQBH and EHT constraints on the angular frequency $\Omega_{\rm H}$, $\angle\beta_2$, and ring diameter $d_{\rm ring}$. 
The green contours cover the LQBH parameters that satisfy $-163^\circ < \angle\beta_2<-127^\circ$, the blue contours represent the angular frequencies of LQBHs that are higher than that of a Kerr BH with $a=0.5$, the solid black contour represents the ring diameter $d_{\rm ring}=42\,\rm \mu as$, and dashed black contours are the $1\,\sigma$ deviation of the $d_{\rm ring}$. Panel A and panel B represent the GRRT result that is blurred and unblurred, respectively.

As shown in Fig.~\ref{fig:M87}, the blurring does not influence the value of $\angle\beta_2$ very much. However, the ring diameters between them are quite different. As discussed in the previous section, the blurred result mostly reveals the size of the inner shadow, which is related to the property of the event horizon \citep{Chael:2021rjo}. The unblurred one shows a stronger constraint on the maximum value of $P\sim0.12$. However, the blurred one shows an upper limit for $P\sim 0.25$. As discussed for Sgr\,A$^*$, we believe blurred images are more reliable. Along with the constraint on $\angle\beta_2$ and $\Omega_{\rm H}$, it provides a smaller possible parameter region in Fig.~\ref{fig:M87}a (see the overlapped region of green and blur contour) with $P\lesssim0.07$ and $0.5\lesssim a\lesssim0.7$.

It is indicated that the information of $\beta_2$ is important to determine the possible parameters for M\,87$^*$ and Sgr\,A$^*$.
With the future polarimetric data of Sgr\,A$^*$, we will be able to make a more accurate constraint on LQG effect of it. In addition to these constraints, the measurement of jet power can offer an additional robust constraint on the parameters of the black hole \citep{Akiyama:2019fyp, EventHorizonTelescope:2022urf}. Calculation of jet power requires explicit GRMHD simulations, which can directly compare the jet power from different BHs and compare it with the EHT constraint. We plan to do such studies in the future.

Considering the uncertainty of the measurement of the BH mass of M~87$^*$, we investigate the impact of different BH masses on our current analysis in Appendix \ref{diff_mass}. In general, the influence on the upper limit for $P$ is small, but smaller BH mass favors relatively larger $P$. In the extreme case ($M_\bullet=5.8\times 10^9\,\rm M_\odot$), no Kerr BH satisfies our constraints on ring diameter.

\section{Summary and discussion}

In this study, using general relativistic radiative transfer (GRRT) simulations, we explored the effect of the polymeric function ($P$) of LQBH on the shadow images. We conducted a range of simulations using the LQBH metric with different values of spin parameters $(a)$ and polymeric function $(P)$. The summaries of our investigation are presented below.

\begin{enumerate}
    \item By performing GRRT calculations with the semi-analytic RIAF model with {\tt ipole}, we found that LQBHs have larger luminous rings than Kerr BHs with the same spin, and the size of the lensing ring and photon sphere increases monotonically with increasing of polymetric function. However, the radius of the event horizon does not follow this trend, which first increases with $P$ and then slightly shrinks when $P$ is close to the allowed maximum value.
    \item  In this study, we used the phase angle of the circular mode of the complex coefficient $\beta_2$ to quantify the polarimetric pattern. We found that for fixed spin $a$, the value of $\angle\beta_2$ increases with increasing polymetric function $P$ when $a\lesssim0.5$ and decreases slightly when $a\gtrsim0.6$. This is because the ratio of radial and angular motion $|u^r/u^\phi|$ at the lensing ring increases and similarly decreases with $P$. Since the polarimetric pattern is influenced by the motion of the accretion flows, as $|u^r/u^\phi|$ changes, the value of $\angle\beta_2$ also evolves with it.
    \item By comparing shadow sizes for different LQBH parameters with EHTC observations, we found that the value of the polymeric function has a bound $P\lesssim0.2$ for Sgr\,A$^*$. This is true if the spin parameter of the Sgr\,A$^*$ is $a\geq0.5$, which is in accordance with \cite{EventHorizonTelescope:2022urf}. However, we also show that the shadow sizes merely cannot constrain these parameters. Other observations, such as jet power and polarization of images, need to be considered simultaneously. For such a study, 3D GRMHD simulations of MAD models are required. We plan to study them in the future.
    \item By the current resolution of EHT, the constraint has some degeneracy between spin $a$ and polymetric function $P$. We cannot distinguish Kerr BH and LQBH only through the ring diameter. Future higher-resolution surveys (e.g., ngEHT) will provide stronger constrain on LQG parameter. 
    \item The measurement of $\angle\beta_2$ in the observation of M\,87$^*$ provide another channel to constrain the $P$ value. Using the same method with Sgr\,A$^*$, we provide a maximum $P$ value for M\,87$^*$ is 0.07, for $0.5\lesssim a\lesssim0.7$. It also implies that we can make stronger constraints for Sgr\,A$^*$ with polarimetric data in the future.

\end{enumerate}

Finally, we would like to discuss some caveats of our study. 
\textbf{The model in this work is not unique within the framework of LQG. In the LQG theory, different quantization schemes lead to black hole solutions with different spacetime properties, e.g. \citep{Modesto:2008im, Gan:2020dkb, Sahu:2015dea, Ashtekar:2018cay, Ashtekar:2018lag, Bojowald:2018xxu}.
Our result in this work implies the property of the metric in \cite{Liu:2020ola}, they used the Newman-Janis method to generate a rotating loop quantum spacetime from a spherically symmetric self-dual black hole solution proposed in \cite{Modesto:2008im}, where the polymeric function $P$ characterizes the loop quantum effect of this spacetime. The self-dual spacetime is derived in the mini-superspace approach based on the polymerization procedure in LQG \cite{Modesto:2008im}. It can be verified that under the transformation $r \rightarrow a_0/r$, where $a_0$ is related to the minimal area gap of LQG, the metric remains invariant. This is achieved with suitable re-parameterization of other variables, thus satisfying the T-duality \cite{Modesto:2008im, Sahu:2015dea,
Ashtekar:2018cay}. We compute the shadow of this rotating LQBH in conjunction with simulations and examine the geometrical properties and polarization information of this shadow. From observations of the shadows of M87 and Sgr\,A$^*$ we find no significant traces of the loop quantum effect. For both of the BHs, the Kerr solution is in the allowed parameter space under the current observation limit. Thus we give an upper limit of $P\lesssim0.2$ for the effect, i.e. if such a self-dual loop quantum effect exists, it should not be larger than this value.} 

Our study only uses a semi-analytical GRRT simulation. The higher-order corrections from the accretion flows are not included (e.g., hot spot, light curve variation, etc.), which we plan to explore in future studies with self-consistent 2D and 3D GRMHD simulations and GRRT calculations. On top of this, studying self-consistent spectral (e.g., image, spectra) and timing properties (e.g., variability) of astrophysical sources demands a two-temperature accretion flow model and different magnetic field configurations \cite[e.g.,][]{Jiang:2023ygs, Dihingia:2022aoc, Mizuno:2021esc}. We will carry forward such study in the future.

\appendix
\section{Geodesic motion in LQBH}\label{sec:geodesic}
	
In this section, we analyze the evolution of massive and massless particles around the rotating LQBH. As we know, a particle follows a geodesic in a given black hole spacetime. To find time-like and null geodesics around the black hole, we can use the Hamilton-Jacobi equation given by,
	\bqn
	\frac{\partial S}{\partial \lambda}=-\frac{1}{2}g^{\mu\nu}\frac{\partial S}{\partial x^\mu}\frac{\partial S}{\partial x^\nu},
	\eqn
where $\lambda$ is the affine parameter and $S$ denotes the Jacobi action of the particle. The Jacobi action $S$ can be separated in the following form,
	\begin{equation}
	S=\frac{1}{2}m^2\lambda-Et+L\phi+S_r(r)+S_\theta(\theta),
	\end{equation}
	where $m$ denotes the mass of the particle moving in the black hole spacetime and $m^2= 1$ or $m^2=0$ for timelike and null geodesics, respectively. $E$ is the energy and $L$ represents the angular momentum of the particle in the direction of the rotation axis. The two functions $S_r(r)$ and $S_\theta(\theta)$ depend only on $r$ and $\theta$, respectively.
	
	Now substituting the Jacobi action into the Hamilton-Jacobi equation, we obtain
	\bqn
	S_r(r)&=&\int^r\frac{\sqrt{R(r)}}{\Delta}dr,\\
	S_\theta(\theta)&=&\int^\theta\sqrt{\Theta(\theta)}d\theta,
	\eqn
	where
	\begin{eqnarray}
    R(r) &=&[X(r)E-aL]^2-\Delta(r)[\mathcal{C}+ m^2 r^2+(L-aE)^2],\\
	\Theta(\theta)&=& \mathcal{C}+(a^2E^2-m^2a^2-L^2\csc^2\theta)\cos^2\theta,
	\end{eqnarray}
	with $ \mathcal{C}$ denoting the Carter constant, where  $\Delta(r)$ is defined by Eq. (\ref{Delta}), $X(r)\equiv k+a^2$.
	
Since the shadow depends on the geodesic equations of photons in the LQBH spacetime. Next, we calculate them explicitly from the from the variations of the Jacobi action, which gives four equations of motion for the evolution of the photon,
	\bqn
	\Sigma\frac{dt}{d\lambda} &=&a(L-aE\sin^2\theta) 
	+\frac{r^2+a^2}{\Delta}[(r^2+a^2)E -aL], \lb{YYY} \\
	\Sigma\frac{d\phi}{d\lambda} &=&\frac{L}{\sin^2\theta}-aE+\frac{a}{\Delta}[(R^2+a^2)E-aL],\\
	\Sigma\frac{dr}{d\lambda} &=&\pm\sqrt{R(r)},\\
	\Sigma\frac{d\theta}{d\lambda}&=&\pm\sqrt{\Theta(\theta)}.\lb{XXX}
	\eqn
where the two choices $\pm$ depend on the sign of the velocity along the angular and radial directions, respectively. The motion of a photon is determined by the two impact parameters, viz.,
	\bqn
	\xi=\frac{L}{E},\qquad \eta=\frac{\mathcal{C}}{E^2}.
	\eqn
To determine the geometric shape of the shadow of the black hole, we need to find the critical circular orbit for the photon, which can be derived from the unstable condition
	\bqn
	R(r)=0,\qquad \frac{dR(r)}{dr}=0,\qquad
\frac{d^2R(r)}{dr^2}>0\lb{ZZZ}
	\eqn
The geometric shape of the shadow is determined by the allowed values of $\xi$ and $\eta$ that fulfill these conditions. In general, the shape of the shadow depends on the rotation parameter $a$ and polymeric function $P$.
	
	
The shadow is a circular disk and it is described by the photon sphere, solving the conditions (\ref{ZZZ}), one finds that, for the spherical motion of photons, the two impact parameters $\xi$ and $\eta$ assume the forms
\begin{align}  \label{AAA}
&\xi=\frac{X_\text{ps}\Delta'_\text{ps}-2\Delta_\text{ps}X'_\text{ps}}{a\Delta'_\text{ps}},\nb\\
&\eta=\frac{4a^2X'^2_\text{ps}\Delta_\text{ps}-\left[\left(X_\text{ps}-a^2\right)   \Delta'_\text{ps}-2X'_\text{ps}\Delta_\text{ps} \right]^2}{a^2\Delta'^2_\text{ps}},
	\end{align}
	where
	
\bqn \label{BBB}
	X(r,M,a,P)&=&a^2+\frac{r^4+a_0}{(2 M P+r)^2},\nb\\
\Delta(r,M,a,P)&=&a^2+\frac{r^2 (2 M-r) \left(2 M P^2-r\right)}{(2 M P+r)^2}.
	\eqn

The parameters $\xi$ and $\eta$ given by Eqs.~(\ref{AAA}) and~(\ref{BBB}) reduce to the expressions of {the} Kerr black hole when  $P=0, a_0=0$. Note that both expressions for ($\xi,\,\eta$) {diverge} as $a\to 0$, however, the expression $\xi^2+\eta$ has a finite value at $a=0$, and it coincides with the spherical static LQBH case \cite{Liu:2020ola}.

\section{GRRT Parameters Survey} \label{Sec:survey}
We implemented a parameter survey using the different values of black hole spin $a$ and polymeric function $P$ listed in table~\ref{list}. We also use multiple inclination angles for the semi-analytical GRRT calculation, which ranges from $5^\circ$ to $30^\circ$.
\begin{table}[]
\centering
\begin{tabular}{@{}ll@{}}
\toprule
a   & P    \\ \midrule
0.2 & 0.00 \\
0.2 & 0.02 \\
0.2 & 0.04 \\
0.2 & 0.06 \\
0.2 & 0.12 \\
0.2 & 0.16 \\
0.2 & 0.24 \\
0.2 & 0.38 \\
0.2 & 0.45 \\
0.3 & 0.00 \\
0.3 & 0.03 \\
0.3 & 0.05 \\
0.3 & 0.06 \\
0.3 & 0.07 \\
0.3 & 0.10 \\
0.3 & 0.15 \\
0.3 & 0.20 \\
0.3 & 0.35 \\
0.4 & 0.00 \\
0.4 & 0.02 \\
0.4 & 0.03 \\
0.4 & 0.05 \\
0.4 & 0.10 \\
0.4 & 0.20 \\
0.4 & 0.25 \\ \bottomrule
\end{tabular}
\begin{tabular}{@{}ll@{}}
\toprule
a   & P     \\ \midrule
0.5 & 0.00  \\
0.5 & 0.04  \\
0.5 & 0.08  \\
0.5 & 0.12  \\
0.5 & 0.16  \\
0.5 & 0.19  \\
0.6 & 0.00  \\
0.6 & 0.02  \\
0.6 & 0.04  \\
0.6 & 0.05  \\
0.6 & 0.09  \\
0.6 & 0.14  \\
0.7 & 0.00  \\
0.7 & 0.03  \\
0.7 & 0.06  \\
0.7 & 0.09  \\
0.8 & 0.00  \\
0.8 & 0.01  \\
0.8 & 0.02  \\
0.8 & 0.03  \\
0.8 & 0.04  \\
0.8 & 0.055 \\
0.9 & 0.00  \\
0.9 & 0.01  \\
0.9 & 0.02  \\ \bottomrule
\end{tabular}
\caption{Black hole parameters used in semi-analytical GRRT calculation.}
\label{list}
\end{table}

\section{Uncertainty on the mass of M~87$^*$}
\label{diff_mass}
\begin{figure}
    \centering
    \includegraphics[height=0.7\linewidth]{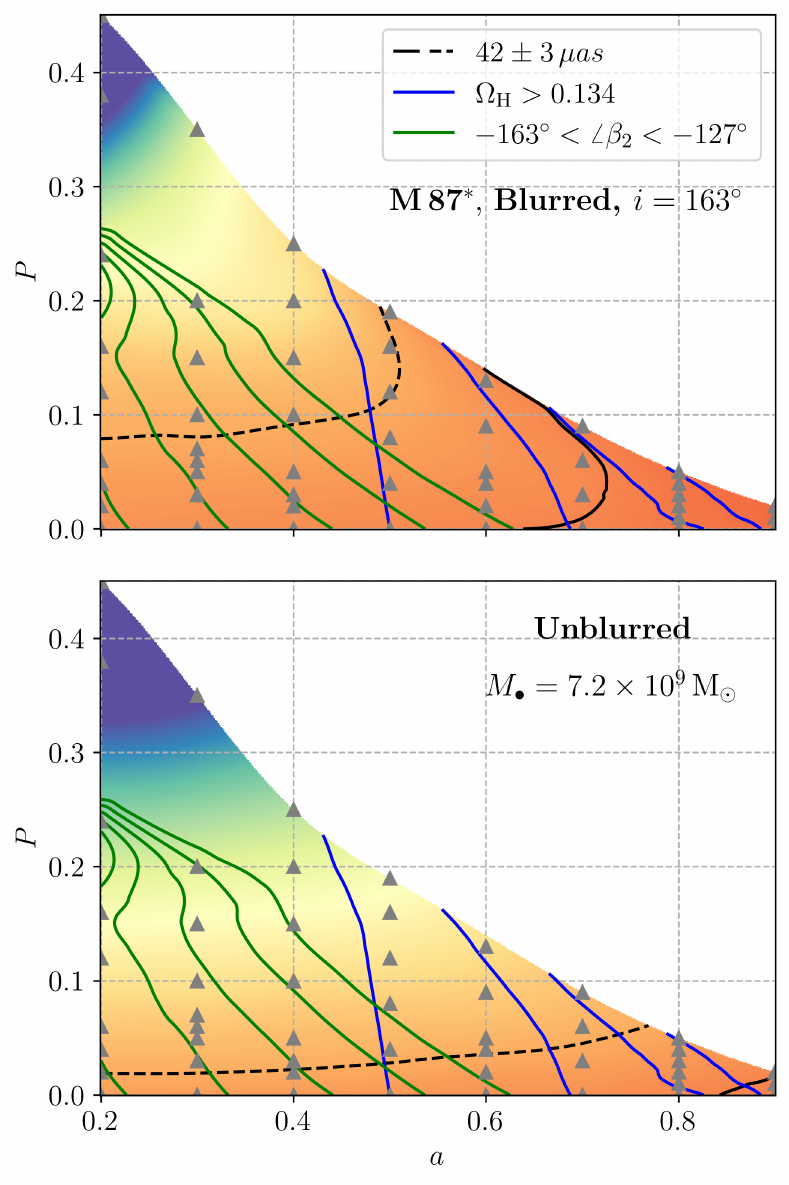}
    \includegraphics[height=0.7\linewidth]{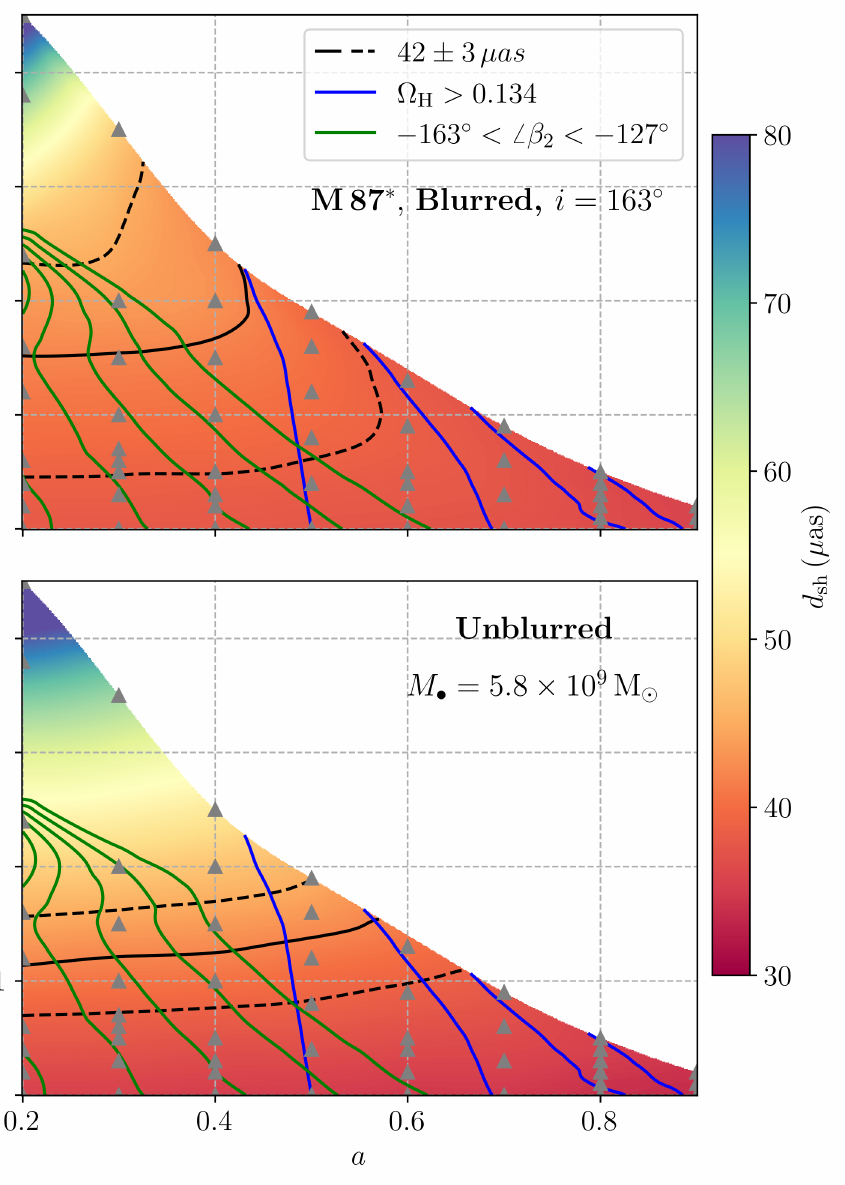}
    \caption{The same with Fig.~\ref{fig:M87} but using the different BH mass, $M_\bullet=7.2\times10^9\,\rm M_{\odot}$ ({\it left}) and $M_\bullet=5.8\times10^9\,\rm M_{\odot}$ ({\it right})
    }
    \label{fig:M87pm}
\end{figure}

The dynamics of the stars around M~87$^*$ are hard to resolve due to the large distance, causing relatively large uncertainty on the mass of the BH. 
From EHT observations of M87$^*$, \citep{EventHorizonTelescope:2019dse}, the mass of central object is estimated as $(6.5\pm0.7) \times 10^9\,\rm M_{\odot}$. Different BH mass will have a direct impact on the ring diameter even if BH parameters ($a$ and $P$) are the same. We performed additional parameter surveys similar to the one in Fig.~\ref{fig:M87} by using two different BH mass which is presented in Fig.~\ref{fig:M87pm}. As shown in this figure, the BH mass does not influence the value of $\angle \beta_2$ very much. $\Omega_{\rm H}$ is not related to BH mass, thus it is not changed. The major difference comes from the constraint of ring diameter. With higher BH mass, the upper limit for $P$ is still $\sim 0.07$. However, as BH mass is lower than $6.5\times10^9\,\rm M_\odot$, the allowed region prefers higher $P$ values. For the extreme case that $M_\bullet=5.8\times10^9\,\rm M_{\odot}$, the $P=0$ cases are ruled out within $1\,\sigma$ deviation of the observation of ring diameter. The upper limit for $P$ is not influenced by BH mass. As a result, the measurement of BH mass is essential for the constraint of LQG.

\acknowledgments
The author gratefully acknowledges insightful discussions with Y. Chen.
This research is supported by the Shanghai Municipality orientation program of Basic Research for International Scientists (Grant No.\,22JC1410600), the National Natural Science Foundation of China (Grant No.\,12273022), and the National Key R\&D Program of China (No.\,2023YFE0101200). 
T.Z. and Q.W. are supported in part by the National Key Research and Development Program of China under Grant No.\,2020YFC2201503, the National Natural Science Foundation of China under Grant No.\,12275238 and No.\,11675143, the Zhejiang Provincial Natural Science Foundation of China under Grant No.\,LR21A050001 and LY20A050002,  and the Fundamental Research Funds for the Provincial Universities of Zhejiang in China under Grant No.\,RF-A2019015.
The simulations were performed on TDLI-Astro, Pi2.0, and Siyuan Mark-I at Shanghai Jiao Tong University.










\end{document}